\documentclass{comjnl}
\usepackage{amsfonts}
\usepackage{amsmath}

\begin{document}

\title[A Scalable Permission Management System With Support of Conditional and Customized Attributes]{A Scalable Permission Management System With Support for Conditional and Customized Attributes}
\author{Baiyu Liu, Abhinav Palia}
\author{Shan-Ho Yang}
\affiliation{Arcules Inc., Irvine, California} \email{\{baiyu , abby, vic\}@arcules.com}

\shortauthors{bli, apl, vya}

\keywords{Permission Management; ABAC; Graph Based Permission Tree; Access Control; Internet of Things}

\begin{abstract}
Along with the classical problem of managing multiple identities, actions, devices, APIs etc. in different businesses, there has been an escalating need for having the capability of flexible attribute based access control~(ABAC) mechanisms. In order to fill this gap, several variations of ABAC model have been proposed such as \textit{Amazon's AWS IAM}, which uses JSON as their underlying storage data structure and adds policies/constraints as fields over the regular ABAC. However, these systems still do not provide the capability to have customized permissions and to perform various operations (such as comparison/aggregation) on them. In this paper, we introduce a string based resource naming strategy that supports the customized and conditional permissions for resource access. Further, we propose the basic architecture of our system which, along with our naming scheme, makes the system scalable, secure, efficient, flexible and customizable. Finally, we present the proof of concept for our algorithm as well as the experimental set up and the future trajectory for this work.
\end{abstract}

\maketitle

\section{Introduction}

There has been a growing need for managing identities, actions, devices,
and third-party APIs in the new era of Internet. Based on the concept 
of role based access control system \cite{sandhu1996role}, researchers proposed 
an attribute based access control (ABAC) model \cite{kuhn2010adding}\cite{hu2015attribute} to provide 
increased flexibility, which has been widely used by the industry. 
IBM Permission Control System \cite{flamini2015attribute} and Amazon's 
AWS IAM \cite{roth2015authorized} are famous derivatives of the ABAC model. There are also third-party 
access control management system that implements ABAC model, such as 

Unlike traditional access 
control systems, with the support of the Internet of Things, conditional 
and customized permissions toward specific resources play an important 
role in modern resource management systems. To support this, Amazon 
introduced identity based policies and resource based policies \cite{awsiampolicies}, where all
the conditional policies are managed in the type of JSON document with
an optional field "Conditions". Inspired by this representation, we 
introduce a string based resource naming strategy that support 
resource, permission, as well as customized attributes introduced by 
permission administrators.

Scalability is another challenge for designing 
a permission management system. Most of the current solutions rely on 
in memory attribute relationships for low latency permission validation.
However, data replication and persistence highly affect the stability of 
the overall system and then create failure to all the calling services. 
In this paper, we use a distributed in-memory graph data model as the 
permission data store which provide fail-safe multi-region data 
existence to keep the scalability and stability.

As an access control system, security should also be maintained using 
certain mechanisms. Goyal et, al. \cite{goyal2006attribute} proposed a 
method that can perform attribute-based encryption for fine-grained 
access control of encrypted data. In our design, we combine the 
attribute-based encryption mechanism with the benefit of using this system
in a private cloud, all permission validation is categorized into two
access layers to ensure both security and usability. 

In the practice of engineering use, access logs with multi-region is also
a key factor of the system design. To maintain the same accessibility as 
the access control data, we choose to use the same data store design
for the two types of data. Weil et, al. \cite{Weil2013auditability} 
introduced a method by applying
role based auditing system to ABAC systems. Inspired by this, we add 
group based logging indexed by permission consumer, which simplifies 
the audit of logging based on attributes without losing the benefit from
ABAC designs.

\section{Architecture}

In our paper, we use an in memory graph data structure as the storage of 
all permission data, and a BigTable \cite{chang2008bigtable} like NoSQL database for persistent permission records. Based on the usage cases,
the system has two trusted levels - permission for binary data and 
permission for encrypted data. Refer to Figure 1.1, Permission 
Enforcement Point is a service that takes in ingress requests and is the
caller of the permission management system (PMS). Action point is hidden from 
the public, protected by the infrastructure firewalls and network settings, handles the actual requests and has the ability to access the database. Usually the permission management system can only be called within the private network (see section 2.1). However for IoT devices and public services outside of the private network, we'll use an encrypted protocol to pass over the action results (see section 2.2). 
\begin{figure}
\centering
\includegraphics[width=3.2in]{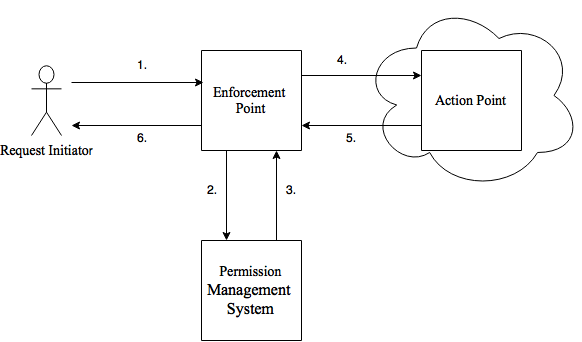}
\caption{Binary Data Permission Flow}\label{fig:binary_perm_flow}
\end{figure}
\subsection{Binary Data Permission Flow}
Binary data permission flow is a widely used architecture for most of permission management systems. As the action point is protected by the infrastructure of private network, data transmitted between enforcement point and the action point is considered as secure so there is no encryption needed.

As shown in FIGURE 1. Binary data is not encrypted and should only be transmitted within a secured private cloud environment.
When doing internal permission grant, the Permission Enforcement Point (PEP) first receives a request to take some actions on other services. Then it checks with the PMS and get back a result of either GRANTED or UNAUTHORIZED. If the result is GRANTED, the PEP performs a call to the action service to continue processing, otherwise the PEP returns UNAUTHORIZED to the caller and stops processing. 

This procedure helps prevent the data from being leaked from the action service without a proper permission being granted by the PMS. On the other hand, this simplifies the architecture that has to be implemented before using the PMS. 

\subsection{Encrypted Data Permission Flow}

When an IoT device or any other public service is the action service, data transferred back to the caller has to be encrypted after the action is taken. Goyal et, al. \cite{goyal2006attribute} gives a way to do permission checks on data exchange with an nonparallel encryption strategy with private key saved at the permission consumer's side. This method solves the case that all the services are exposed to the public network, where the data flow is not protected by the features of a private network. 

As shown in FIGURE 2. The permission system issues a short time public key to the action point to encrypt the data. Then the consumer gets the encrypted data from the action point and uses the private key stored to decrypt the data. For this case, there are no enforcement points needed. Action point can directly talk to the permission system to check for permissions.
\begin{figure}
\centering
\includegraphics[width=3.2in]{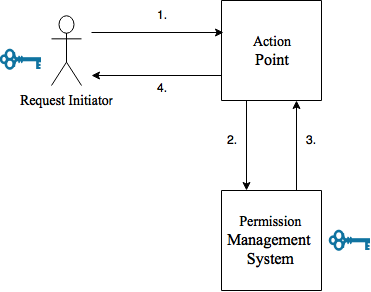}
\caption{Encrypted Data Permission Flow}\label{fig:binary_perm_flow}
\end{figure}

\section{Permission Graph Model} \label{Model}
We use a directed graph structure to implement our algorithm. As shown in FIGURE 3, we define the permission graph $G$ to be
\begin{align*}
G = \{(\mathbb{C}, N, E(n_1, n_2))\ |\ n_1, n_2 \in N, n_2 \not\in \mathbb{C},\nonumber \\
cycle(N, E)\ is\ false\} \nonumber \\ \\
and \ cycle(N, E)\ is\ true\ if 
\exists m \in \mathbb{N}\ s.t. \nonumber \\
E(n_1, n_2), E(n_2, n_3), ... , E(n_m, n_1)\ are\ in\ G \nonumber
\end{align*}
where $N$ is a collection of graph node $n$ represented by a resource name $R$ where a set of permissions $P$ is associated, $E(n_1, n_2)$ is a set of directed edges representing connections between two nodes $n_1$ and $n_2$, and $\mathbb{C}$ is a collection of consumers where no direct edges point to. Then we can easily get $\mathbb{C} \subset N$ from the definition. In this graph, we call $n_2$ the parent of $n_1$ if $E(n_1, n_2)$ is in $G$. And intuitively, we call $n_1$ the child of $n_2$ under the same condition. To better describe and model the permission structure, inspired by Amazon \cite{bhatt2017access}, we use a resource name to represent an item, a permission, or a conditional permission within the system. Accordingly, there are a total of three types of resource names: item, permission and conditional permission. We give the definition of a resource name $R$ as a 3, 4 or 5-tuple according to the type of resource name:
\begin{flalign*} 
R_{item} = (b, i, s) \\
R_{perm} = (b, i, s, l) \\
R_{cond} = (b, i, s, l, v) \nonumber
\end{flalign*}
\begin{figure}
\centering
\includegraphics[width=3.2in]{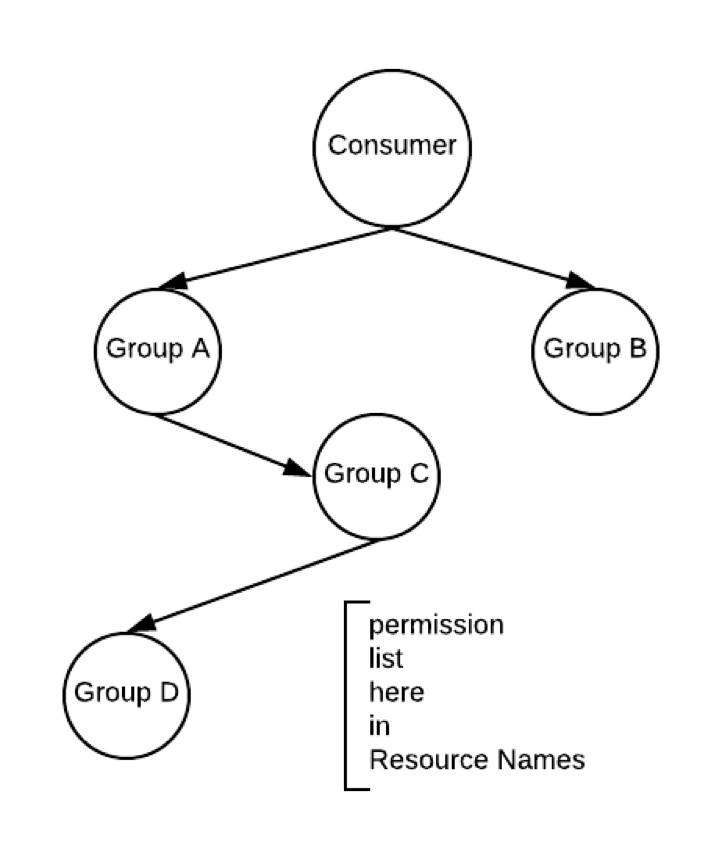}
\caption{Permission Graph Model}\label{fig:graph_model}
\end{figure}
where $b$ is base string, $i$ is identifier, $s$ is attribute or scope, $l$ is permission level and $v$ is permission condition value. To simplify the levels for a given permission, we require $l \in \{admin, edit, view\}$ where the system user can give definition and privilege regarding the three levels. Value $v$, in this definition, has to be comparable for a given attribute or scope $s$ to suffice the calculation requirements. We now define the notion of a conflict between two feature resource names and then present the two core operations (\textit{unite}, $\sqcup$ and \textit{overwrite}, $\sqcap$) required to aggregate the feature list for an entity.
\par Conflict function $C$ on two ARN vectors $\alpha = (b_\alpha, i_\alpha, s_\alpha, l_\alpha, v_\alpha)$ and $\beta = (b_\beta, i_\beta, s_\beta, l_\beta, v_\beta)$ is defined as

\begin{equation}
C(\alpha, \beta): ( b_\alpha = b_\beta, i_\alpha = i_\beta, s_\alpha = s_\beta, l_\alpha \neq l_\beta, v_\alpha \neq v_\beta) \nonumber 
\end{equation}

We can further classify $C$ as $C_l(\alpha,\beta)$, conflict due to different level and $C_v(\alpha,\beta)$, conflict due to different range value with $l_\alpha = l_\beta$. Similarly, \textit{non-conflict} function $\tilde{C}$ is described- 

\begin{equation}
\tilde{C}(\alpha, \beta): ( b_\alpha = b_\beta, i_\alpha = i_\beta, s_\alpha = s_\beta, l_\alpha = l_\beta, v_\alpha = v_\beta) \nonumber
\end{equation}

Since each node is modeled as an entity having a list of resource name. Let $A$ and $B$ denote two peer nodes such that $A= \{\alpha_1, \alpha_2, ...\}$ and $B = \{\beta_1, \beta_2, ...\}$, for $C_l$, we define-
\begin{equation}
A \sqcup B = (A \cup B)\mid _{\tilde{C}(\alpha,\beta)} \cup (Max._{(l)}\{A,B\})\mid _{C_l(\alpha,\beta)}\nonumber \displaybreak[0] 
\end{equation}
Or in case of $C_v$, we can write-
\begin{equation}
(A \cup B)\mid _{\tilde{C}(\alpha,\beta)} \{Max._{(v)}(A,B)\}\mid _{C_v(\alpha,\beta)} \nonumber
\end{equation}

Since we keep the maximum of the level ($C_l$) or the maximum range value ($C_v$) while performing \textit{unite} operation, for the sake of convinience, we can write a generalized form of the above equation which we will be using for further proofs-

\begin{equation}
A \sqcup B = (A \cup B)\mid _{\tilde{C}(\alpha,\beta)} \cup \{Max.(A,B)\}\mid _{C(\alpha,\beta)} \displaybreak[0] 
\end{equation}

Next, we define the \textit{overwrite} operation for two nodes $A$ and $B$ such that node $B$ is the child of node $A$-

\begin{equation}
A \sqcap B = (A \cup B)\mid _{\tilde{C}} \cup
(A)\mid _{C} \displaybreak[0] 
\end{equation}


Using the above definitions, we can show that \textit{unite} operation follows the associative law. 
\par \textbf{Lemma 1}:  $P \sqcup (Q\sqcup R) = (P \sqcup Q) \sqcup R$
However, for \textit{overwrite} operation, the order of operation matters and therefore, it does not follow the associative law. Also, these operations follow usual distributive laws with regular set operations.  
\par \textbf{Lemma 2}: $P \sqcup (Q \cup R) = (P \sqcup Q) \cup (P \sqcup R)$ 
\par \textbf{Lemma 3}: $P \sqcap (Q \cup R) = (P \sqcap Q) \cup (P \sqcap R)$ 
\par We prove these lemmas in appendix A.

\par For the simplest graph with source node $P$ having children $Q$ and $R$ ($Q$ and $R$ being peers), the output set of aggregate algorithm on node $P$ denoted by $\hat{P}$ is defined as-

\begin{equation}
\hat{P} = P \sqcap  (Q \sqcup R) 
\end{equation}
We can now prove the validity of the distributive law over these operations.
\begin{flushleft}
\textbf{Lemma 4}: $P \sqcap (Q \sqcup R) = (P \sqcap Q) \sqcup (P \sqcap R)$
\par\textit{Proof}: The equation on the right can be expanded as-
\end{flushleft}
\begin{align*}
&P \sqcap ((Q \cup R)\mid _{\tilde{C}} \cup \{Max.(Q,R)\}\mid _{C}) \nonumber 
\end{align*}
From \textit{Lemma} 3 we can write-
\begin{equation}
(P \sqcap (Q \cup R)\mid_{\tilde{C}_(q,r)}) \cup (P \sqcap \{Max.(Q,R)\}\mid _{C_(q,r)})
\nonumber 
\end{equation}
Expanding the above equation-
\begin{align*}
&[(P \cup Q \cup R)\mid_{\tilde{C}_(p,q,r)} \cup (P) \mid_{C(p,q,r)}] \cup 
[\{(P\cup Q)\mid_{\tilde{C}(p,q), q>r}  \nonumber \\
&\quad\cup (P)\mid_{C(p,q)}\} \cup \{(P\cup R)\mid_{\tilde{C}(p,r), q<r} \cup (P)\mid_{C(p,r)}\}] \nonumber \\ \\
&[(P \cup Q)\mid_{\tilde{C}(p,q)} \cup (P)\mid_{C(p,q)})]\cup [(P \cup R)\mid_{\tilde{C}(p,r)} \cup (P)\mid_{C(p,r)})] \nonumber \\
&\quad\cup [Max.(P\cup Q)\mid_{C(p,q)}, (P\cup R)\mid_{C(p,r)}] \nonumber
\end{align*}
which, by definition is-
\begin{align*}
&(P \sqcap Q) \sqcup (P \sqcap R)
\end{align*}
And by induction, it is easy to prove for $n$ child nodes. Therefore,
\begin{equation}
P \sqcap (Q_1 \sqcup Q_2 \sqcup... Q_n) = (P \sqcap Q_1) \sqcup (P \sqcap Q_2) \sqcup ... (P \sqcap Q_n) 
\end{equation}

This aggregate function is applied on each node in the graph connected to the root in order to get its feature list. We suggest a bottom up approach which involves doing a \textit{depth-first} search on the nodes and then aggregating features for each node in a \textit{breadth first} manner.  
\par Using the above implementation strategy, the complexity of the aggregate algorithm can be calculated by defining $n(C)$ as the number of conflicts in a graph having a total of $N$ nodes. If each node on an average has $\overline m$ features, intuitively, the worst case occurs when $n(C) = N.\overline m$, and therefore the average complexity can be written as $\Theta(N.\overline m)$. 

\section{Future Work}
In this paper, we presented our strategy to have a customizable ABAC system which uses our \textit{aggregate} algorithm over the graph derived from the underlying database. As our future trajectory, we propose to use an in memory graph database to store each item or frequently used items. This will reduce the overhead of constructing a graph every time a request is made to access it. Further, in the graph database, it will be easier to to propagate the changes to all the connected nodes. 
\par In order to increase the robustness, we propose to have a master - slave model where both the master as well as the slaves are identical having a data layer (graph database) and a compute layer (\textit{aggregate} algorithm). However, only the master stays inside the memory and is updated frequently whereas the slave nodes are redundant nodes which will be updated after regular intervals and will replace the master in case of a failure. This architecture can be further extended by having multiple replicas. The number of layers can be decided based on the criticality of the business. 
\par For bigger organizations with a global outreach, we propose to have a multi-regional support on distributed databases, where data sharing among these data centers can be done using global messaging queues. It would also require the master to perform additional operations including routing of data based on the markers and the location. 

\section{Conclusions} \label{Conclusions}

\appendix
\section{}
\begin{flushleft}
\textbf{Lemma 1}:  $P \sqcup (Q\sqcup R) = (P \sqcup Q) \sqcup R$
\end{flushleft}
\begin{align*}
&\textit{Proof:} \quad (P \cup [(Q \cup R)]\mid_{\tilde{C}}) \cup (Max.\{P, (Max.\{Q,R\}\})\mid_{C}) \nonumber \\
&([P\cup Q]\cup R\mid_{\tilde{C}}) \cup (Max.\{Max.\{P,Q\}, R\}\mid_{C}) \nonumber
\end{align*}
which is $(P\sqcup Q) \sqcup R$

\begin{flushleft}
\textbf{Lemma 2}: $P \sqcup (Q \cup R) = (P \sqcup Q) \cup (P \sqcup R)$ 
\end{flushleft}
\begin{align*}
&\textit{Proof:}\quad ([P \cup (Q \cup R)]\mid_{\tilde{C}}) \cup (Max.\{P, (Q\cup R)\}\mid_{C}) \nonumber \\
&([P\cup Q] \cup [P\cup R])\mid_{\tilde{C}} \cup (Max.\{Max.\{P, Q\},Max.\{P, R\}\mid_{C}) \nonumber 
\end{align*}
which is equal to $(P \sqcup Q) \cup (P \sqcup R)$, since it doesn't matter how we calculate the maximum of all the sets.

\begin{flushleft}
\textbf{Lemma 3}: $P \sqcap (Q \cup R) = (P \sqcap Q) \cup (P \sqcap R)$ 
\end{flushleft}
\begin{align*}
&\textit{Proof:}\quad ([P \cup (Q \cup R)]\mid_{\tilde{C}}) \cup (Max.\{P, (Q\cup R)\}\mid_{C}) \nonumber \\
&([P \cup Q] \cup [P\cup R])\mid_{\tilde{C}} \cup (P)\mid_{C}\nonumber 
\end{align*}
which can be combined to form $(P \sqcap Q) \cup (P \sqcap R)$

\nocite{*}

\bibliographystyle{compj}
\bibliography{reference}

\end{document}